\documentclass{article}
\usepackage[dvips]{epsfig}
\usepackage{amsfonts}
\usepackage{amssymb}
\usepackage{latexsym}
\usepackage{amsmath}
\usepackage{makeidx}
\DeclareGraphicsExtensions{.jpg, .eps}
\newcommand{\newc}{\newcommand}

\newc{\ra}{\rightarrow}

\newc{\lra}{\leftrightarrow}
\newcommand{\dis}{\displaystyle}
\def\ket#1{|{#1}\rangle}

\newc{\beq}{\begin{equation}}
\newc{\eeq}{\end{equation}}
\newc{\barr}{\begin{eqnarray}}
\newc{\earr}{\end{eqnarray}}

\newcommand{\bvec}[1]{\mbox{\boldmath$#1$}}

\newcommand{\bvecb}[1]{\scriptsize \mbox{\boldmath$#1$}}

\begin{document}
\begin{titlepage}

%\markboth{G. S. Kliros and
% P. C. Divari} {Localized wavefunctions and magnetic band structure for
% lateral semiconductor superlattices.}
%
\title{\large  \bf LOCALIZED WAVEFUNCTIONS AND MAGNETIC BAND
 STRUCTURE FOR LATERAL SEMICONDUCTOR SUPERLATTICES }
%
%
%
%\author{G. S. KLIROS\footnote{ Hellenic Air-force Academy,
%Dekelia Air-Force Base, Attica TGA-1010, Greece}}
%
%\address{Department of Electronic and Communication Engineering, HellenicAir-force Academy,
%Dekelia Air-Force Base, Attica TGA-1010, Greece\\
%gsksma@hol.gr  }
%
%\author{P. C. DIVARI}
%\address{Department of Physics, University of Ioannina, GR-45110 Ioannina,
%Greece\\}
%
%\maketitle
%
%\begin{history}
%\received{Day Month Year} \revised{Day Month Year}
%\end{history}

\author{G. S. Kliros$^{a}$ and P.C. Divari$^{b}$ \\
{\small\it $^{a}$Department of Electronic and Communication
Engineering }\\
{\small\it Hellenic Air-force Academy,  Dekelia Air-Force Base, Attica TGA-1010, Greece}\\
{\small\it $^{b}$Department of Physics, University of Ioannina,
GR-45110 Ioannina, Greece}}

%\date{\small\it Received 4 July 2005; revised 1 November 2005;
%accepted 8 December 2005}
%%Available online 27 December 2005}

\maketitle

\begin{abstract}
In this paper we present calculations on the electronic band
structure of a two-dimensional lateral superlattice subject to a
perpendicular magnetic field by employing a projection operator
technique based on the ray-group of magnetotranslation operators.
We construct a new basis of appropriately symmetrized Bloch-like
wavefunctions as linear combination of well-localized
magnetic-Wannier functions. The magnetic field was consistently
included in the Wannier functions defined in terms of
free-electron eigenfunctions in the presence of external magnetic
field in the symmetric gauge. Using the above basis, we calculate
the magnetic energy spectrum of electrons in a lateral
superlattice with bi-directional weak electrostatic modulation.
Both a square lattice and a triangular one are considered as
special cases. Our approach based on group theory handles the
cases of integer and rational magnetic fluxes in a uniform way and
the provided basis could be convenient for further both analytic
and numerical calculations.
\end{abstract}

%\keywords{magnetic translation group; band structure; lateral
%superlattice, modulated 2DES.}
\end{titlepage}

\section{Introduction}

Lateral semiconductor superlattices are realized by modern
lithographic techniques to impose a periodic pattern onto a
two-dimensional electron system (2DES) at a semiconductor
heterostructure. This pattern can be a uni- or bi-directional
periodic potential modulation of different strength allowing for a
variety of artificial periodic structures.\cite{Peet,Weis}
Magnetotransport  measurements in such structures have revealed
novel oscillations in the conductivity, the Weiss oscillations
which are superimposed on top of the well-known Shubnikov-de Haas
oscillations.\cite{Deut} Besides Weiss oscillations (a
commensurability effect),novel magneto-resistance oscillations
with $1/B-$periodicity have been observed in short-period
superlattices due to magnetic miniband energy
spectrum.\cite{CAlbr} The above developments have renewed the
interest in the problem of Bloch electrons in a magnetic field,
which was extensively studied in the past by renowned
theorists\cite{Peie} but it is experimentally accessible now with
these artificial periodic structures.\cite{Albr1}

The first theoretical approach to the problem is due to
Peierls\cite{Peie} and it is based  on the formulation of an
effective single-band Hamiltonian, arising from a tight-binding
dispersion relation $E(\bvec{k})$, through the substitution
$E(\bvec{k}+e\bvec{A}/\hbar)$, where $\bvec{A}$ the vector
potential. When combined to a semiclassical picture\cite{Onsa},
the effective Hamiltonian theory leads to the idea of quantized
magnetic orbits, which can be commensurate with the lattice
period. Within the framework of effective single-band Hamiltonian
theory, further investigations lead to butterfly-like patterns for
the energy spectrum known as Hofstadter spectrum.\cite{Hofs,Clar}
This spectrum reflects, as a consequence of Harper$'$s
equation\cite{Harp}, the splitting of the single Bloch-band into
magnetic subbands, according to the number of magnetic flux quanta
piercing the unit cell of the lattice. Some recent experimental
results\cite{Albr} supports the existence of the Hofstadter$'$s
spectrum in high magnetic field regime. Furthermore, an
experimental realization of the Hofstadter's spectrum was achieved
considering the transmission of microwaves\cite{Kuhl} and acoustic
waves\cite{Rich} through an array of scatterers.

In recent years special attention has been paid to investigate the
energy miniband structure\cite{Gran}as well as the transport
properties of two-dimensional short-period lateral supperlattices
in the presence of a perpendicular magnetic field $\bvec{B}$ in
the case of weak modulation.\cite{Ross,Lang}  Some calculations of
superlattice electronic structure under the above conditions have
used perturbation theory\cite{Usov,Peet1} or a tight-binding
scheme\cite{Labb} and look for solutions of the one-electron
Hamiltonian that are linear combination of the unperturbed
extended Landau wavefunctions. Another approach to the problem is
based on Ferrari's wavefunctions\cite{Silb}, that is,
wavefunctions constructed out of coherent states associated to a
von Neumann-type lattice.\cite{Ferr} Besides the effect of
electric modulation, the effect of magnetic modulation on the
superlattice energy spectrum has been studied\cite{Shi,Chang} and
a formal connection between these two has been
investigated.\cite{Niu}

In this paper we present calculations on the electronic band
structure of a lateral superlattice subject to a perpendicular
magnetic field by employing a projection operator technique in the
context of magnetic translation group (M.T.G) theory. We use a new
set of magnetic-consistent Wannier functions given in terms of
Schrauben functions introduced by Jannussis\cite{Jann} in order to
construct Bloch-like eigenfunctions that reflect the crystal
symmetry of the superlattice. Schrauben functions describe free
electrons in a uniform magnetic field in the symmetric gauge and
have been used in the past for magnetoconductance
calculations.\cite{Chun,Klir}

The paper has the following structure: In Section 2 the model is
formulated and the projection operator of the M.T.G corresponding
to the problem is introduced. In Section 3 we construct a basis of
appropriately symmetric Bloch-like wavefunctions as linear
combination of Schrauben functions. In Section 4, the electronic
spectrum of the 2DES, under weak periodic modulation, is obtained.
In Section 5, we present specific results for periodic modulations
with rectangular or triangular symmetry. The density of states is
calculated as well. We conclude with a summarizing Section 6.

\section{Projection operator of magnetic translation group}

To describe the electrons in the conduction band of a lateral
superlattice at the AlGaAs-GaAs interface in a constant
perpendicular magnetic field $\bvec{B}=B\hat{z}$, we employ a
model of strictly two-dimensional electron gas (2DEG) with a 2D
periodic potential modulation $V(\bvec{r})$,
$\hspace{3pt}\bvec{r}\equiv(x,y)$.

Following the common practice, we adopt an effective mass
approximation, to take into account the effect of the crystalline
atomic structure over the charge carriers. The system is described
by the well-known single-particle Hamiltonian of Bloch electrons
in a magnetic field: \beq
\label{eq.2.1}\mathcal{H}=\frac{1}{2m^\ast} \big
(\bvec{p}+e\bvec{A} \big)^2 +V(\bvec{r}) \eeq where we neglect the
Zeeman splitting and spin-orbit interactions which are very small
for GaAs systems.

The Hamiltonian (\ref{eq.2.1}) does not commute with the usual
translations operators \beq \label{eq.2.2}
\hat{T}(\bvec{R})=\exp{\big(-\frac{i}{\hbar}\bvec{R}\cdot
\bvec{p}\big)} \eeq where $ \bvec{R}$ is any translation vector,
but commutes with unitary operators of the form \beq
\label{eq.2.3}
\hat{T}_m(\bvec{R})=\exp{\bigg[-\frac{i}{\hbar}\bvec{R}\cdot(\bvec{p}-e\bvec{A})}\bigg]
\eeq introduced by Brown\cite{Brow} and Zak\cite{Zak}, if the
vector potential $\bvec{A}$ fulfills the condition \beq
\label{eq.2.4} \frac{\partial A_j}{\partial x_k}+\frac{\partial
A_k}{\partial x_j}=0 \hspace{15pt} \mbox{for}
\hspace{10pt}j,k=1,2,3 \eeq This relation holds, for the symmetric
gauge $ \bvec{A}(\bvec{r})=\frac{1}{2}(\bvec{B}\times \bvec{r}) $,
which was used by both authors. It is worth noting that
introducing local gauge one can consider any vector
potential.\cite{Wieg,Flor}

The above operators (\ref{eq.2.3}) form the magnetic translation
group (M.T.G). This is in fact a ray group\cite{Brow1}, because
the product of two elements of the M.T.G yields another element,
multiplied by a constant phase
 \beq \label{eq.2.5}
 \hat{T}_m(\bvec{R}_1) \hspace{3pt} \hat{T}_m(\bvec{R}_2)   =
  \hat{T}_m(\bvec{R}_1+\bvec{R}_2)
\exp{\bigg [-\frac{i e}{\hbar}\bvec{R}_2\cdot
\bvec{A}(\bvec{R}_1)\bigg ]} \eeq where $ \bvec{R}_1$ and $
 \bvec{R}_2 $ are  translation  vectors.
 As a consequence of (\ref{eq.2.5}), we can verify the commutation relation \beq
\label{eq.2.6}
 \Big [ \hat{T}_m(\bvec{R}_1), \hspace{3pt} \hat{T}_m(\bvec{R}_2) \Big ]  =
  2 i \sin \bigg [ -\frac{ e}{2\hbar}
\bvec{B}\cdot \big (\bvec{R}_1\times \bvec{R}_2 \big )
 \bigg ]
 \hat{T}_m(\bvec{R}_1+\bvec{R}_2)
\eeq where the symmetric gauge for the vector potential has been
used.

Therefore, the magnetic translation operators commute with the
Hamiltonian, but in general not with each other. Only in the case
 that the phase constant in (\ref{eq.2.6}) is a multiple of $\pi$,
\beq
 \label{eq.2.7}
 \frac{e}{2\hbar}\bvec{B}\cdot \big (\bvec{R}_1\times \bvec{R}_2 \big
 )=\pi q , \hspace{10pt} q \in \Bbb N
\eeq we get
 $\Big [ \hat{T}_m(\bvec{R}_1), \hspace{3pt}
\hat{T}_m(\bvec{R}_2) \Big ]=0 $ and the operators form an Abelian
group. The condition (\ref{eq.2.7}) is compatible with the demand
that the eigenfunctions of the system satisfy the Born-von Karman
periodic boundary conditions, e.g. are periodic under magnetic
translations corresponding to the full lattice size.

The term $\bvec{B}\cdot \big (\bvec{R}_1\times \bvec{R}_2 \big )$
in (\ref{eq.2.7}) is the magnetic flux through the cell spanned by
lattice vectors $\bvec{R}_1 $ and $ \bvec{R}_2$ and can be written
as an integer multiple ($p$) of the flux per unit cell
$\Phi=\Phi_0\bvec{B}\cdot \big (\bvec{a}_1\times \bvec{a}_2 \big
)$, where $\Phi_0=e/\hbar$ is the magnetic flux quantum.

Consequently, the condition(\ref{eq.2.7}) is fulfilled each time
the number of magnetic flux quanta through the unit cell is a
rational number: \beq
 \label{eq.2.8}
 \frac{\Phi}{\Phi_0}=\frac{\hspace{4pt}q}{\hspace{-1pt}p},
 \quad \mbox{with} \quad q,p \quad \mbox{mutually prime integers}
 \eeq thus, imposing a commensurability relation between the
magnetic field intensity and the lattice size.

 After a brief preamble on the M.T.G, we follow Brown$'$s
formulation\cite{Zak} to introduce the projection operators of
M.T.G. We consider the irreducible representations of the M.T.G
labelled by the magnetic crystal momentum vector $\bvec{q}$,
restricted to a single magnetic Brillouin Zone (MBZ), or to
$1/p\times 1/p $ part of it if $p\ne 1 $. The $\bvec{q}-th$
irreducible representation is given by a familiar relation \beq
 \label{eq.2.9}
D^{\bvecb{q}}(\bvec{R})=\exp{\big(-i\bvec{q}\cdot\bvec{R}}\big)
\hspace{3pt}D^{0}(\bvec{R})
 \eeq
where $\bvec{R}=n_1\bvec{a}_1+n_2\bvec{a}_2$, with  $\bvec{a}_1$,
$\bvec{a}_2$ the primitive lattice vectors and
$\bvec{q}=q_1\bvec{b}_1+q_2\bvec{b}_2$ with  $\bvec{b}_1$,
$\bvec{b}_2$ are the primitive reciprocal-lattice vectors. The
vectors
  $\bvec{a}_{1,2}$ and  $\bvec{b}_{1,2}$ are related in the usual
  fashion $\bvec{a}_i \cdot \bvec{b}_j=2\pi \delta_{ij}$ and
  $n_{1,2}$, $q_{1,2}$ are integer indices.

  The $p\times p$ matrix $D^{0}(\bvec{R})$ for any translation
  $\bvec{R}$ is generated from those corresponding to the
  primitive lattice vectors
\beq
 \label{eq.2.10}
D^{0}_{kk'}(\bvec{a}_1)=\delta_{kk'} \exp{\bigg [ 2\pi i
(k-1)\frac{q}{p}\bigg ]}, \quad
D^{0}_{kk'}(\bvec{a}_2)=\delta_{k,k'-1}^{mod \hspace {1pt} p}
 \eeq
and the ray-group multiplication law (\ref{eq.2.5}) yields to \beq
 \label{eq.2.11}
D^{0}_{kk'}(\bvec{R})=\delta_{k,k'-n_2}^{mod \hspace {1pt} p}
\exp{\bigg [ i\pi \frac{q}{p} n_1\Big ( n_2+2(k-1)\Big )\bigg ]},
\quad \mbox{with} \qquad k,k'=0,1,2,\cdots p-1
 \eeq
where the modulo Kronecker delta, $\delta_{ij}^{mod \hspace {1pt}
p}$ equals 1 if $i$ and $j$ are equal or differ by a multiple of
$p$ and vanishes otherwise.

Using (\ref{eq.2.9}) and (\ref{eq.2.11}) we obtain the following
form for the irreducible representations
 \begin{equation}
 \label{eq.2.12}
D^{(q_1,q_2)}_{kk'}(\bvec{R})=\delta_{k,k'-n_2 }^{mod \hspace
{1pt} p}
 \exp{ \bigg [ -2\pi i \Big(
n_1q_1+ n_2q_2\Big ) \bigg ] } \exp \bigg [ i\pi \frac{q}{p}n_1
\Big ( n_2+2(k-1) \Big ) \bigg ] \end{equation} We can see that
increasing any of $q_1$, $q_2$ by $1/p$ produces an equivalent
representation. Thus the values of $\bvec{q}$ have to be
restricted to a single  magnetic Brillouin Zone: $0\leq
q_1,q_2<1/p$. The rank of matrices implies the existence of
$p$-partner functions transforming according to different rows of
the same representation $D^{(q_1,q_2)}_{kk'}$

We introduce the projection operator of M.T.G. of Bloch electrons
in a uniform magnetic field\cite{Altm,Fisc}
 \beq
 \label{eq.2.13}
P^{ \bvecb{q}}_{kk'}(\bvec{R})=\sum_{\bvecb{R}}\Big
[D^{\bvecb{q}}_{kk'}(\bvec{R}) \big]^{\ast}\hat{T}_m(\bvec{R})
 \eeq
 which projects out the component belonging to $k'$-th row of the
 representation $\bvec{q}$ and then generates the partner
 belonging to $k$-th row. Using the expression for the magnetic
 translation operator (\ref{eq.2.3}), Eq. (\ref{eq.2.13}) leads to
\beq
 \label{eq.2.14}
P^{\bvecb{q}}_{kk'}(\bvec{R})=\sum_{\bvecb{R}}
C_{\mbox{\scriptsize $\bvec{R}$ }}
\hspace{3pt}\exp{\Big(i\bvec{q}\cdot\bvec{R} \big)}\exp
\bigg[\frac{i }{\hbar} \bvec{R}\cdot \big (\bvec{p}- \frac{e
}{2\hbar} \bvec{B}\times \bvec{R} \big )
 \bigg]
 \eeq
 where
\beq
 \label{eq.2.15}
C_{\mbox{\scriptsize $\bvec{R}$ }}=\delta_{k,k'-n_2 }^{mod \hspace
{1pt} p}\exp{\bigg [ -i\pi \frac{q}{p} n_1\Big (n_2+2(k-1)\Big
)\bigg ]}
 \eeq

 We finally remark that acting the projection operator $
 P^{\bvecb{q}}(\bvec{R})$ to an arbitrary function we can find a
 partner function of the $\bvec{q}$-th representation of M.T.G. In
 the next section, we find a basis of Bloch electrons in a
 magnetic field by acting the found projection operator
 (\ref{eq.2.14}) onto a basis of free electrons in a magnetic
 field.

\section{Localized wave-functions of Bloch electrons in a magnetic
field}

Landau functions, eigenfunctions of angular momentum and localized
functions based on coherent states\cite{Ferr,Hutc} are usually
used to describe free electrons moving in a plane under the
influence of a magnetic field. The coherent states\cite{Feld} on a
von Neumann lattice are the most localized eigenfunctions for an
electron in magnetic field. For such a lattice, with a single flux
quantum per unit cell, the area of the unit cell equals $2\pi
l^2$, where $l= ( \hbar/eB )^{1/2}$ is the magnetic length.

Jannussis\cite{Jann} working in the symmetric gauge, introduced a
different basis of wave functions for free electrons in uniform
magnetic field, who called them Schrauben functions: \beq
 \label{eq.3.1}
%\mbox{\tt \Large{u}}
\psi_{n,\bvecb{k}}(\bvec{r})=C_n\exp{\bigg[-(\kappa_x^2+\kappa_y^2)l^2+i\bvec{k}
\cdot \bvec{r}\bigg]}\big( -\kappa_y-i\kappa_x \big)^n
 \eeq
 where $n\in\Bbb N$ is the Landau index, $\bvec{k}$ the electron
 wavevector and $\kappa_{x}=k_{x}+y/2l^2$, \hspace{3pt}
 $\kappa_{y}=k_{y}-x/2l^2$,\hspace{2pt}The normalization coefficient is given by
\beq\label{eq.3.2}
 C_n=\sqrt{\frac{1}{2\pi l^2}\frac{(2l^2)^n}{n!}}
 \eeq
 The above functions have several important properties, some of
 which are:

 i)They form a basis in the Hilbert space of free electrons in a
 uniform magnetic field.
 \beq
 \label{eq.3.3}
\sum_{n,\bvecb{k}} %\mbox{\tt \Large{u}}
\psi_{n,\bvecb{k}}^{\ast}(\bvec{r}) %\mbox{\tt \Large{u}}
\psi_{n,\bvecb{k}}(\bvec{r'})=\delta(\bvec{r}-\bvec{r'})
 \eeq
 \beq
 \label{eq.3.4}
\bigg (
%\mbox{\tt \Large{u}}
\psi_{n,\bvecb{k}}(\bvec{r}), %\mbox{\tt \Large{u}}
\psi_{n',\bvecb{k}}(\bvec{r}) \bigg
 )=\delta_{nn'}
 \eeq

 ii)They are fully localized around the centers
\beq
 \label{eq.3.5}
\bvec{r}_m(x_m,y_m) \hspace{3pt} :  \hspace{5pt} x_m=l^2k_y,
\hspace{5pt} y_m=-l^2k_x
 \eeq
 and in the two-dimensional plane take the symmetric form :
\begin{align}
 \label{eq.3.6}
%\mbox{\tt \Large{u}}
\psi_{n,\bvecb{k}}(\bvec{r})= & C_n\bigg [(x-x_m)^2-i(y-y_m)^2
\bigg ]^n
\nonumber \\
 &\cdot \exp{\bigg [ -\frac{1}{4l^2}\Big
[(x-x_m)^2+(y-y_m)^2+\frac{i}{2l^2}(x_my-xy_m) \Big ] \bigg ] }\;
 \end{align}
 Also, the probability density $\rho=\psi^{\ast}\psi$
 has a rotational symmetry about the centers $(x_m,y_m)$ and
takes the form of the Poisson distribution \beq
 \label{eq.3.7}
\rho(r_m)=\frac{1}{2\pi l^2}\frac{\big (r_m^2/2l^2\big
)^n}{n!}\exp{\bigg ( -\frac{r_m^2}{2l^2}\bigg )}
 \eeq
 where $r_m^2=(x-x_m)^2+(y-y_m)^2$

 iii) They obey Harper$'$s symmetric condition\cite{Harp}, i.e.
 for a lattice translation $\bvec{R} $
\beq
 \label{eq.3.8}
%\mbox{\tt \Large{u}}
\psi_{n,\bvecb{k}}(\bvec{r}+\bvec{R})=\exp{\bigg[\frac{ie}{\hbar}\bvec{A}(\bvec{R})\cdot\bvec{r}
+i\bvec{k} \cdot \bvec{r}\bigg]}% \mbox{\tt \Large{u}}
\psi_{n,\bvec{\kappa}}(\bvec{r}) \quad , \hspace{15pt}
\bvec{\kappa}=\bvec{k}-\frac{e}{\hbar}\bvec{A}(\bvec{R})
 \eeq

 It has been shown\cite{Jann1} that a Wannier function $w_n(\bvec{r}-\bvec{r}_m) $ for electrons in a
 magnetic field can be obtained directly form the Schrauben
 function if the wave-vector $\bvec{k} $ is substituted by the
 vector potential $\displaystyle{-e\bvec{A}(\bvec{r}_m)/\hbar} $

 In what follows, we construct a basis for two-dimensional Bloch
 electrons in a uniform magnetic field applying the projection
 operator method described in Section 2. In order to find the
 principal partner function we choose the projection operator
 $P^{\bvecb{q}}_{11}(\bvec{R})$ (see Eq. (\ref{eq.2.14})),
 projecting onto the first row of $\bvec{q}$-th representation.

 Acting the projection operator Eq. (\ref{eq.2.14}) onto the
 functions $\psi_{n,0}(\bvec{r})$ of the basis  (\ref{eq.3.1}) we find the
 following partner functions of the $\bvec{q}$-th representation of the M.T.G.
\begin{align}
 \label{eq.3.9}
%\hspace{-15pt}
\Psi_{n,\bvecb{q}}(\bvec{r})=&\sum_{\bvecb{R}}C_{\bvecb{R}}
\hspace{3pt}\exp{  \Big( i\bvec{q}\cdot\bvec{R} \big)
}\psi_{n,\bvecb{k}_0}(\bvec{r}) \nonumber \\
=&\sum_{\bvecb{R}}C_{\bvecb{R}} \hspace{3pt}\exp{ \Big(
i\bvec{q}\cdot\bvec{R}  \big) }w_{n}(\bvec{r}-\bvec{R}),\quad
\bvec{k}_0=-\frac{e}{\hbar}\bvec{A}(\bvec{R})\;
 \end{align}
 where $w_{n}(\bvec{r}-\bvec{R})$ are the magnetic Wannier
 functions of free electrons in a uniform magnetic field\cite{Jann1},
\begin{align}
 \label{eq.3.9a}
w_{n}(\bvec{r}-\bvec{R})=& \hspace{3pt}C_n\bigg \{ \frac{1}{2l^2}[(x-n_1a_1)+i(y-n_2a_2)] \bigg \}^n  \nonumber \\
&\cdot \exp \bigg \{ -\frac{1}{4l^2}\Big [
(x-n_1a_1)^2+(y-n_2a_2)^2 \Big ] +\frac{i}{2l^2}(yn_1a_1-xn_2a_2)
\bigg \}
\end{align}
 where $a_1$, $a_2$ are the 2D-lattice periods along $x$ and $y$
 directions respectively and the normalization constant $C_n$ is given by Eq.(\ref{eq.3.2}).

 Therefore, Eq. (\ref{eq.3.9}) represents modified Bloch functions
 which are formally expanded in the corresponding magnetic Wannier
  functions, that is, functions  strongly localized on the lattice
  sites. Note that, for commensurate magnetic fields
  (\ref{eq.2.7}) one can recover Bloch$'$s theorem
\beq
 \label{eq.3.10}
\hat{T}(\bvec{a})\Psi_{n,\bvecb{q}}(\bvec{r})=\exp{\big (
-i\bvec{q}\cdot\bvec{a} \big )}\Psi_{n,\bvecb{q}}(\bvec{r})
%\mbox{\small $\bvec{R}$ }
  \eeq
  with the values of the magnetic crystal momentum
  $\bvec{q}=(q_x,q_y)$ restricted to the first magnetic Brillouin
  zone.
Also, $\Psi_n$$_{,\bvecb{q}} $ with different values of $\bvec{q}$
are orthogonal since they belong to different irreducible
representations of M.T.G.
  Finally, after some straightforward steps, we arrive at
  the following expression for the eigenfunctions Eq. (\ref{eq.3.9})
\begin{align}
\label{eq.3.11}
\Psi_{n,\bvecb{q}}(\bvec{r})= &\sum_{n_1n_2}C_{n_1n_2}
C_n\hspace{3pt}
 \exp{ \bigg [ i (q_xn_1a_1+q_yn_2a_2) \bigg ]} \bigg \{ \frac{1}{2l^2}[(x-n_1a_1)+i(y-n_2a_2)] \bigg \}^n  \nonumber \\
&\cdot \exp \bigg \{ -\frac{1}{4l^2}\Big [
(x-n_1a_1)^2+(y-n_2a_2)^2 \Big ] +\frac{i}{2l^2}(yn_1a_1-xn_2a_2)
\bigg \}
  \;
\end{align}
where $C_{n_1n_2}= \exp{( -i\pi n_1n_2q/p )}$ with $n_2$ integer
multiple of $p$.
 The values of crystal momentum $\bvec{q}=(q_x,q_y)$  are
 determined from the periodic boundary conditions for
 $\Psi_{n,\bvecb{q}}(\bvec{r})$ with respect to magnetic
 translations with periods $a_1$ and $a_2$.

 Figure 1 shows the contour plots of the modulus
 $|\Psi_{0,\bvecb{q}}(\bvec{r})|$ for the
 case of a square periodic lattice $(a_1=a_2=100nm)$ in the
 presence of a perpendicular magnetic field that corresponds to two flux
 quanta per unit cell, for different values of the crystal momentum
 $\bvec{q}$ in the first Brillouin  zone. It is seen that the
 wavefunctions possess the full symmetry of the lattice and have regularly
 distributed zeros as it is expected from the translational invariance of the system.

\section{Electron energy spectrum under weak periodic modulation potential}

Now we can find the energy spectrum of Bloch electrons in a
constant magnetic field and weak periodic modulation potential
$V(\bvec{r})=V(\bvec{r}+\bvec{R})$ using the basis $\big
\{\ket{n,\bvec{q}} \big \}$ introduced in the preceding Section 3.

It is well-known\cite{Jann} that the matrix elements of the first
part of the Hamiltonian (\ref{eq.2.1}) which describes free
electrons in a uniform magnetic field are given by \beq
 \label{eq.4.1}
\Big  \langle {n,\bvec{q}} \Big \vert {\frac{1}{2m^\ast} \big
(\bvec{p}+e\bvec{A} \big)^2} \Big \vert  {n',\bvec{q}} \Big
\rangle=(n+\frac{1}{2}\hspace{3pt})
\hbar\hspace{1pt}\omega_c\hspace{2pt}\delta_{nn'}
%
% \me{n,\bvec{q}}{\frac{1}{2m^\ast} \big (\bvec{p}+e\bvec{A}
%\big)^2}{n',\bvec{q}}
 \eeq
 where $\omega_c=eB/m^\ast$ is the cyclotron frequency and $n$
 labels the Landau levels.

 In order to find the corresponding energy spectrum of Bloch
 electrons we first evaluate the matrix elements of the periodic
 potential. Since $V(\bvec{r}) $ is periodic with respect to $\bvec{R}$
 it can be expressed in terms of a Fourier series
\beq
 \label{eq.4.2}
V(\bvec{r})=\sum_{\bvecb{G}} V_{\bvecb{G}}\exp{\Big (
i\bvec{G}\cdot \bvec{r} \big )} \eeq with the Fourier coefficients
given by \beq
 \label{eq.4.3}
V_{\mbox{\scriptsize $\bvec{G}$ }}=\frac{1}{A}\int_{A}
d^{\hspace{2pt}2}r V(\bvec{r}) \exp{\big ( -i\bvec{G}\cdot\bvec{r}
\big )}
 \eeq
where $A$ stands for the area of the unit cell in the
two-dimensional lattice and $ \bvec{G}$ is a reciprocal lattice
vector.

In the two dimensional lattice $(x,y)$ the potential of Eq.
(\ref{eq.4.2}) is written as \beq
 \label{eq.4.4}
V(x,y)=\sum_{g_1g_2}V_{g_1,g_2}    \exp{ \bigg[i\Big
(\frac{2\pi}{a_1}g_1x+\frac{2\pi}{a_2}g_2y \Big )\bigg ]}
 \eeq
with $g_1$, $g_2$ integers and $a_1$, $a_2$ the modulation periods
in $x$ and $y$ direction, respectively.

Using the basis (\ref{eq.3.11}), the matrix elements of the
potential $ V(x,y)$ can be derived for each crystal momentum
$\bvec{q}=(q_x,q_y) $:
\begin{align}
 \label{eq.4.5}
V_{nn'}\equiv  &\Big  \langle {n,\bvec{q}} \Big \vert V(x,y) \Big
\vert
{n',\bvec{q}}\Big  \rangle=\sum_{g_1g_2}\sum_{n_1n_2}\sum_{n_1'n_2'} C_{n_1n_2}^{\ast}C_{n_1'n_2'}V_{g_1g_2}\nonumber \\
\cdot& \ \exp \Big [iq_xa_1 (n_1-n_1')+iq_ya_2(n_2-n_2') \Big ]\nonumber \\
\cdot& \exp \bigg [ -\frac{a_2^2}{l^2}
(n_2-n_2')(n_2-n_2'+2g_1\frac{\Phi_0}{\Phi}) -\frac{a_1^2}{l^2}
(n_1-n_1') (n_1-n_1'-2g_2 \frac{\Phi_0}{\Phi}) \bigg ]\nonumber \\
\cdot& \exp \bigg [ i\pi\Big
[(n_1'n_2-n_2'n_1)\frac{\Phi}{\Phi_0}+(n_1+n_1')g_1+(n_2+n_2')g_2\Big]\bigg
]I_{nn'}(g_1,g_2)
  \;
\end{align}
where
\begin{align}
 \label{eq.4.6a}
 I_{nn'}(g_1,g_2)=&(-1)^{n+n'}\sqrt{\frac{n'!}{n!}}\hspace{3pt}
\Bigg ( \sqrt{\frac{a_1}{a_2}}g_2+i \sqrt{\frac{a_2}{a_1}}g_1
\Bigg )^{n-n'}\Bigg ( \pi \frac{\Phi_0}{\Phi} \Bigg)^{ {(n-n')}/{2}} \nonumber \\
\cdot& \exp{( -u/2 )}\hspace{2pt} L_{n'}^{n-n'}(u) , \qquad
\mbox{for} \quad n\geq n'
  \;
\end{align}
 and
\begin{align}
 \label{eq.4.6b}
 I_{nn'}(g_1,g_2)=&(-1)^{n+n'}\sqrt{\frac{n!}{n'!}}\hspace{3pt}
\Bigg ( -\sqrt{\frac{a_1}{a_2}}g_2+i \sqrt{\frac{a_2}{a_1}}g_1
\Bigg )^{n'-n}\Bigg ( \pi \frac{\Phi_0}{\Phi} \Bigg)^{ {(n'-n)}/{2}} \nonumber \\
\cdot& \exp{( -u/2 )}\hspace{2pt}L_{n}^{n'-n}(u) , \qquad n'\geq n
  \;
 \end{align}
with $ \dis{u=\pi( \frac{a_2}{a_1}g_1^2+\frac{a_1}{a_2}g_2^2
)\frac{\Phi_0}{\Phi}}$ and $L_m^n(u)$ the associate Laguerre
polynomials.

Furthermore, from the expression (\ref{eq.4.5}) one sees that for
high magnetic fields or if the lattice points are far apart each
other, the matrix elements $V_{nn'}$ approach zero unless the
following conditions are valid:
\begin{align}
 \label{eq.4.7}
n_1-n_1'=2g_2\frac{\Phi_0}{\Phi} \qquad \mbox{and} \qquad
n_2-n_2'=-2g_1\frac{\Phi_0}{\Phi}
  \;
 %&L^S, L^V \longrightarrow 1 \nonumber \\
 %&L^P \longrightarrow 0 \nonumber \;  \, .
 \end{align}
Under the above conditions, which mean that the magnetic flux
through a unit cell of the periodic lattice should be a rational
number, the matrix elements (\ref{eq.4.5}) take the form \beq
 \label{eq.4.8}
V_{nn'}=\sum_{g_1g_2}V_{g_1g_2}
  \exp \Big [2i\frac{\Phi_0}{\Phi}(g_2a_1q_x-g_1a_2q_y) \Big ]
I_{nn'}(g_1,g_2)
 \eeq

Finally, the first order energy perturbation can be found by
calculating the diagonal matrix elements from Eqs. (\ref{eq.4.8})
and (\ref{eq.4.6a}): \beq
 \label{eq.4.9}
V_{nn}(q_x,q_y)=\sum_{g_1g_2}V_{g_1g_2}
  \exp \Big [2i\frac{\Phi_0}{\Phi}(g_2a_1q_x-g_1a_2q_y) \Big ]
\exp{ ( -u/2)}L_{n}(u)
 \eeq
Thus, the complete energy spectrum to first order approximation
can be written as \beq
 \label{eq.4.10}
E_n(q_x,q_y)=(n+\frac{1}{2}\hspace{3pt})
\hbar\hspace{1pt}\omega_c\hspace{2pt}+V_{nn}(q_x,q_y)
 \eeq

\section{Energy band structure for lateral superlattices with
weak modulation}

In this section, we present analytical results concerning the
energy spectrum of superlattices with perfect rectangular or
triangular symmetry applying the formalism of the preceding
Section 4. We consider smooth periodic potentials described by a
few lowest Fourier components with weak modulation strengths.

It is well-known that the numerical solution of the corresponding
tight-binding equation\cite{Wang} for 2D-lattices in perpendicular
magnetic field, shows that each Landau level splits into several
sub-bands when a rational number of flux quanta $\Phi_0$ pierces
the unit cell and that the corresponding gaps are exponentially
small. In what follows, we assume that the gaps are closed due to
disorder in samples of not exceptionally high mobilities, that is,
we assume that the Landau levels are bands neglecting the above
fine structure. This assumption is justified since the reported
experiments\cite{Chow}did not indicate that this fine structure of
the energy spectrum was resolved.

\subsection{\bf Lateral superlattices with rectangular
   symmetry}

We consider a two-dimensional periodic modulation with rectangular
symmetry, i.e: \beq
 \label{eq.5.1}
V(x,y)=V_x\cos{(\frac{2\pi}{a_1}x)} +V_y\cos{(\frac{2\pi}{a_2}y)}
 \eeq
where $V_x$, $V_y$ are the modulation strengths and $a_1$ $a_2$
the periods in the corresponding directions. We assume that the
modulation is weak compared to the cyclotron energy $\hbar
\omega_c$ so that Eq. (\ref{eq.4.10}) is valid and mixing of
different Landau levels can be neglected.

In terms of the plane wave expansion Eq. (\ref{eq.4.4}), the
periodic potential (\ref{eq.5.1}) can be described by four Fourier
coefficients
\begin{equation}
V_{(\pm 1 ,0)}=V_x/2 \qquad \mbox{and} \qquad V_{(0 ,\pm
1)}=V_y/2\nonumber \end{equation}
 Applying Eq. (\ref{eq.4.10}) we find the
energy spectrum
\begin{align}
 \label{eq.5.2}
 E(q_x,q_y)= (n+\frac{1}{2}\hspace{3pt})
\hbar\hspace{1pt}\omega_c\hspace{2pt}+V_xF(u_x)\cos{(\frac{2a_1}{\pi}u_xq_y)}
 +&V_yF(u_y)\cos{(\frac{2a_2}{\pi}u_yq_x)}
  \;
 \end{align}
 where
 \beq F(u)=\exp{(-u/2)}L_n(u), \quad
 u_x=\frac{a_2}{a_1}(\pi\Phi_0/\Phi)\quad \mbox{and} \quad
 u_y=\frac{a_1}{a_2}(\pi\Phi_0/\Phi)\nonumber
 \eeq
%are \hspace{3pt}  $''$inverse magnetic flux$''$ parameters.

Thus, the unperturbed Landau levels broaden into bands, with width
equal to $2(V_x|F_n(u_x)|+V_y|F_n(u_y)|)$, that oscillates with
inverse magnetic flux $\Phi_0/\Phi$ . Using the asymptotic form of
Laguerre polynomials in the large n-limit, it can be shown that
the bandwidth tends to zero when $2\sqrt{n\pi}=(\lambda +1/4)\pi$
and reaches maximum when $2\sqrt{n\pi}=(\lambda -1/4)\pi$ with
$\lambda$ an  integer. Therefore, the energy bandwidths in Eq.
(\ref{eq.5.2}) oscillate with the Landau index. One important
consequence of the non-zero bandwidth is finite
electron-velocities leading to a finite band
conductivity\cite{Vasi} which is  absent when the modulation is
not present.

In Fig.2 we plot the energy spectrum for the first two Landau
levels, $n=0,1$, \hspace{2pt}modulation strengths  $V_x=V_y=0.5$
meV and for (a) an integer $(\Phi/\Phi_0=2)$ and (b) a rational
$(\Phi/\Phi_0=5/4)$ number of flux quanta. The square superlattice
period has been taken $a=100$ nm. It is seen that the energy
spectrum reflects the symmetry of modulated square lattice.

The interesting characteristics in the energy spectrum of the weak
modulated superlattice are also reflected in the density of states
(DOS) per unit area  defined by \beq
 \label{eq.5.3}
g(E)=\frac{g_s}{2\pi l^2}\sum_{n=0}^\infty \hspace{3pt}
\sum_{q_xq_y} \delta[E-E_n(q_x,q_y)]\eeq where $g_s=2$ is the
electro-spin degeneracy. For the spectrum given by Eq.
(\ref{eq.5.2}) the DOS per unit area  takes the following form:
\begin{align}
\label{eq.5.4} g(E)=&g_0\sum_{n=0}^\infty \int_0^{2\pi}d\theta \{
[V_xF_n(u_x)]^2-[E-(n+1/2\hspace{3pt})
\hbar\hspace{1pt}\omega_c-V_yF_n(u_y)\cos{\theta}]^2 \}^{-1/2}
\end{align}
where $g_0=1/(\pi^3l^2)$.  Figure 3 plots the DOS per unit area
for the same values of parameters as in Fig.2. As can be seen, the
DOS does not exhibit Van Hove singularities at the edges of each
Landau band reflecting the 2D nature of the electron motion in
this band with finite mean velocities. Finally, we have to note
that similar results for the band structure of a two dimensional
surface superlattice have been obtained previously by direct
application of perturbation theory to a different unperturbed
basis of extended Landau wavefunctions.\cite{Peet1,Wang}

\subsection{\bf Lateral superlattices with triangular
    symmetry}

We now consider a  laterally modulated  2DEG  with triangular
symmetry, i.e:
\beq
 \label{eq.5.5}
V(x,y)=V_x\cos{(\frac{2\pi}{a_1}x)}\cos{(\frac{2\pi}{a_2}y)}
+V_y\cos{(\frac{4\pi}{a_2}y)}
 \eeq
where  $a_1=a$ and $a_2=\sqrt{3}a$. This periodic modulation can
be realized in a GaAs/AlGaAs heterostructure by means of a diblock
copolymer nanolithography technique.\cite{Meli}

The above periodic potential can be described by six Fourier
coefficients
\begin{equation}
V_{(1,\pm 1 )}=V_{(-1,\pm 1 )}=V_x/4 \qquad \mbox{and} \qquad
V_{(0 ,\pm 2)}=V_y/2\nonumber \end{equation} where we assume again
the condition of weak modulation $V_x,V_y<<\hbar \omega_c$.
 Applying Eq.
(\ref{eq.4.10}) we find the energy spectrum
\begin{align}
 \label{eq.5.6}
 E(q_x,q_y)= &(n+1/2\hspace{3pt})
\hbar\hspace{1pt}\omega_c\hspace{2pt}+V_xF(u_x+u_y)\cos{(\frac{2a}{\pi}u_xq_y)}\cos{(\frac{2\sqrt{3}a}{\pi}u_yq_x)}\nonumber
\\
 +&V_yF(4u_y)\cos{(\frac{4\sqrt{3}a}{\pi}u_yq_x)}
  \;
 \end{align}
 where the parameters $u_x$, $u_y$ and the function $F_n(u)$ are
 defined in the preceding subsection 5.1.

 As in the rectangular case, the Landau levels have broadened
 into bands with a bandwidth equal to
 $2(V_x+V_y)|F_n(4u_y)$, that oscillates with inverse magnetic
 flux $\Phi_0/\Phi $ and Landau band index n. In Fig.4 the energy bands
 for $n=0$ and $n=1$, are plotted  for the same values of parameters as
 in the rectangular case. The  energy spectrum reflects the symmetry of
 the modulated triangular lattice.

Furthermore, the DOS per unit area of the superlattice  with
triangular symmetry
 can be calculated according to Eq. (\ref{eq.5.3}) and is given  by
\begin{align}
\label{eq.5.7} g(E)=&g_0\sum_{n=0}^\infty
\int_0^{2\pi}d\theta \{ [V_xF_n(u_x+u_y)\cos{\theta}]^2\nonumber\\
&-[E-(n+1/2\hspace{3pt})
\hbar\hspace{1pt}\omega_c-V_yF_n(4u_y)\cos{2\theta}]^2 \}^{-1/2}
\end{align}
where $g_0=1/(\pi^3l^2)$. The DOS for this case has similar
characteristics to the one shown in Fig.3 and is not plotted.

\subsection{\bf  Exact energy spectrum }

The approximate analytic energy spectrum  of the preceding
subsections $5.1$ and $5.2$, provides a good insight into the
problem and can facilitate the study of other properties, e.g.,
the thermodynamic and the transport properties of the
superlattice. On the other hand, the exact energy spectrum can be
obtained by diagonalizing numerically the Hamiltonian matrix
produced by means of Eq.(\ref{eq.4.8}). The diagonalization is
carried out by truncating the matrix dimensions while including a
sufficient number of Landau levels. We were careful to choose the
rank of the matrix being diagonalized large enough to ensure the
convergence of our result. In Fig.5 the exact energy spectrum of
the superlattice with square symmetry, given by Eq.(\ref{eq.5.2})
and the approximate one are plotted as a function of the crystal
momentum $(q_x,q_y=0)$ for the first ten Landau levels, where the
modulation strengths are $V_x=V_y=0.5$ meV and $\Phi/\Phi_0=2$.

We see that the approximate energy spectrum (solid curve) agrees
well in the whole range of crystal momentum with the exact one
(dotted line). Similar agreement is found for other values of
magnetic flux $\Phi/\Phi_0$ provided that the condition of weak
modulation is satisfied.

\section{\bf Concluding Remarks}

In summary, we have studied the energy spectrum of electrons in
lateral semiconductor superlattices subject to a perpendicular
magnetic field. Such systems, in which the simultaneous influence
of a bi-directional periodic potential and a magnetic field is
studied, are realized nowadays in layered semiconductor structures
with modern technologies of growth and lithography.

We introduced a projection operator method applied to irreducible
representations of the magnetotranslation group in order to
construct an appropriately symmetrized basis of Bloch-type
wavefunctions in terms of a new set of magnetic-consistent Wannier
functions. The well-localized Wannier functions in this linear
combination, were produced by eigenfunctions of free electrons in
a magnetic field obtained in the symmetric gauge.

Using the above Bloch-type wavefunctions  we have calculated the
magnetic energy spectrum of Bloch-electrons in a two dimensional
periodic potential obtaining similar results to previous
pertubative or numerical approaches. The spectrum is expressed as
a function of crystal-momenta, defined in a magnetic Brillouin
zone, which are good quantum numbers of the system. Especially, we
have presented explicit results concerning the band structure of
short-period lateral superlattices with both square and triangular
symmetry. The obtained oscillatory behavior of energy bandwidths,
in the weak modulation regime, manifest themselves in the
magnetoresistance measurements where they give rise to the so
called Weiss oscillations in the dependance of the conductance on
the magnetic field strength.

Finally, we note that our approach provides a symmetry-adapted
basis-functions  based on group theory and handles the cases of
integer and rational magnetic flux in a uniform way. For rational
magnetic fluxes it is assumed that the fine structure of Landau
bands can be neglected in samples of not exceptionally high
mobilities, since the mini-gaps in a Landau band are closed due to
disorder. From the presented results, it seems that our basis is
appropriate in the intermediate regime of magnetic fields where
both the lattice period and the magnetic length are comparable in
magnitude. We also expect that this basis is convenient for
further both analytic and numerical calculations.

\bigskip
\clearpage
%\section*{References}

\clearpage

 \begin{figure}[htb]
%\hspace{2cm}
\begin{center}
\includegraphics[width=1.0\textwidth]{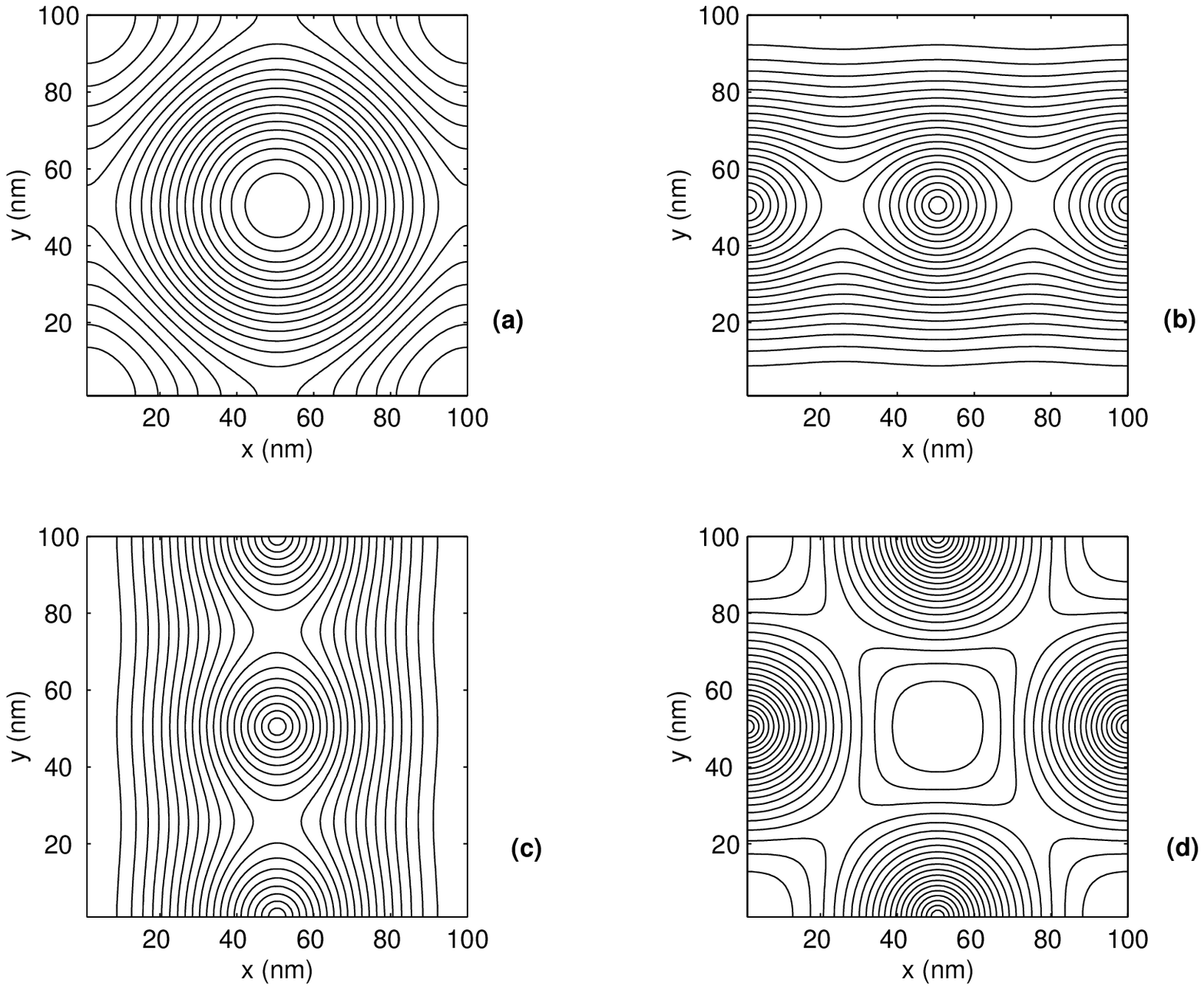}\\
\vspace*{8pt} \caption{Contour plots of the modulus
$|\Psi_{0,\bvecb{q}}(\bvec{r})|$ for a square lattice with $a=100$
nm and $\Phi/\Phi_0=2$,
 for different values of the crystal momentum $\bvec{q}$ in the first Brillouin
 zone: (a)\hspace{2pt} $(q_x,q_y)=(0,0)$, \hspace{2pt} (b)
 \hspace{3pt}$(q_x,q_y)=(\pi/a,0)$, \hspace{2pt}(c)
 \hspace{3pt}$(q_x,q_y)=(0,\pi/a)$, \hspace{2pt}(d)
 \hspace{3pt}$(q_x,q_y)=(\pi/a,\pi/a)$}
\label{ss3.f1}
\end{center}
\end{figure}
\clearpage
\begin{figure}[htb]
%\vspace{1cm}
\includegraphics [width=0.55\textwidth,bb= 45 45 530 530]{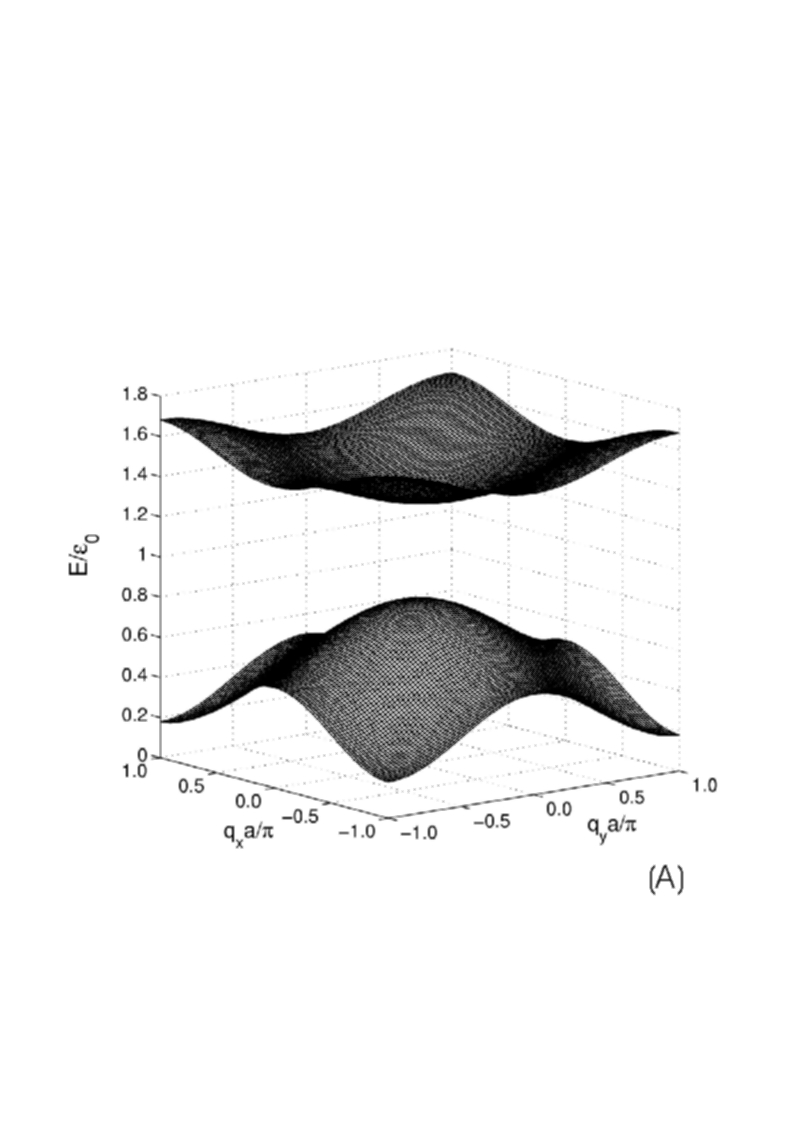}
\includegraphics[width=0.55\textwidth, bb= 45 45 530 530]{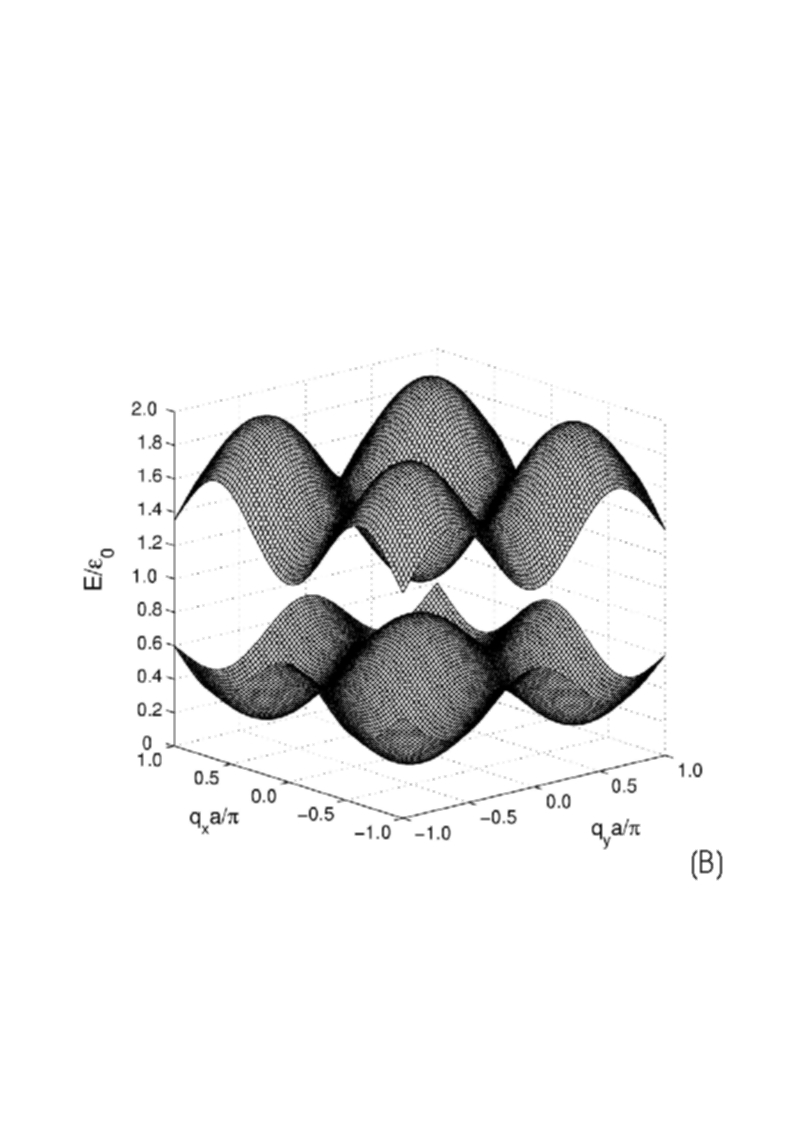}

\caption{Magnetic energy bands $n=0,1$, of the square lattice as a
function of the crystal momentum $(q_x,q_y)$ for (A)\hspace{1pt}
$\Phi/\Phi_0=2 $. and (B) for $\Phi/\Phi_0=5/4 $.The modulation
strengths are \hspace{1pt} $V_x=V_y=0.5$ meV and the energy
eigenvalues are plotted in units of $\epsilon_0=\hbar\omega$.}
\label{fig:2}
\end{figure}

\clearpage
\begin{figure}[htb]
\centerline{\psfig{file=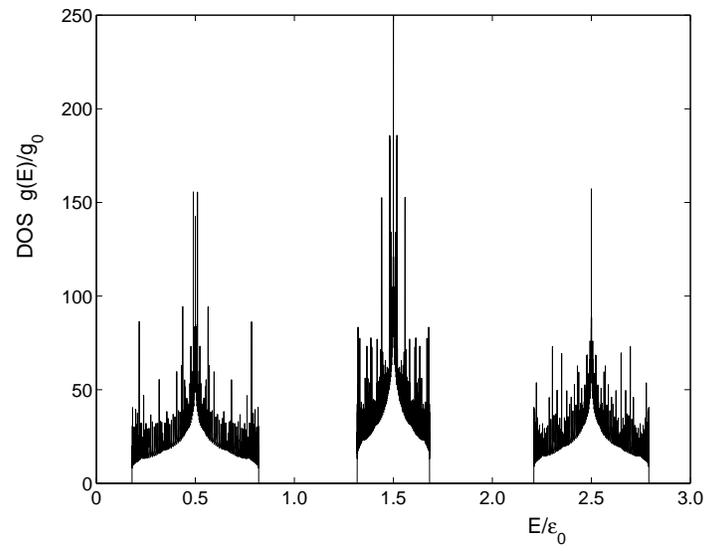,width=3.65in}} \vspace*{8pt}
\caption{Density of states  per unit area of the square lattice
for the same values of parameters as in Fig.2. } \label{ss3.f3}
%\end{center}
\end{figure}

\clearpage
\begin{figure}[htb]
%\vspace{1cm}
\includegraphics [width=0.55\textwidth ,bb= 45 45 530 530]{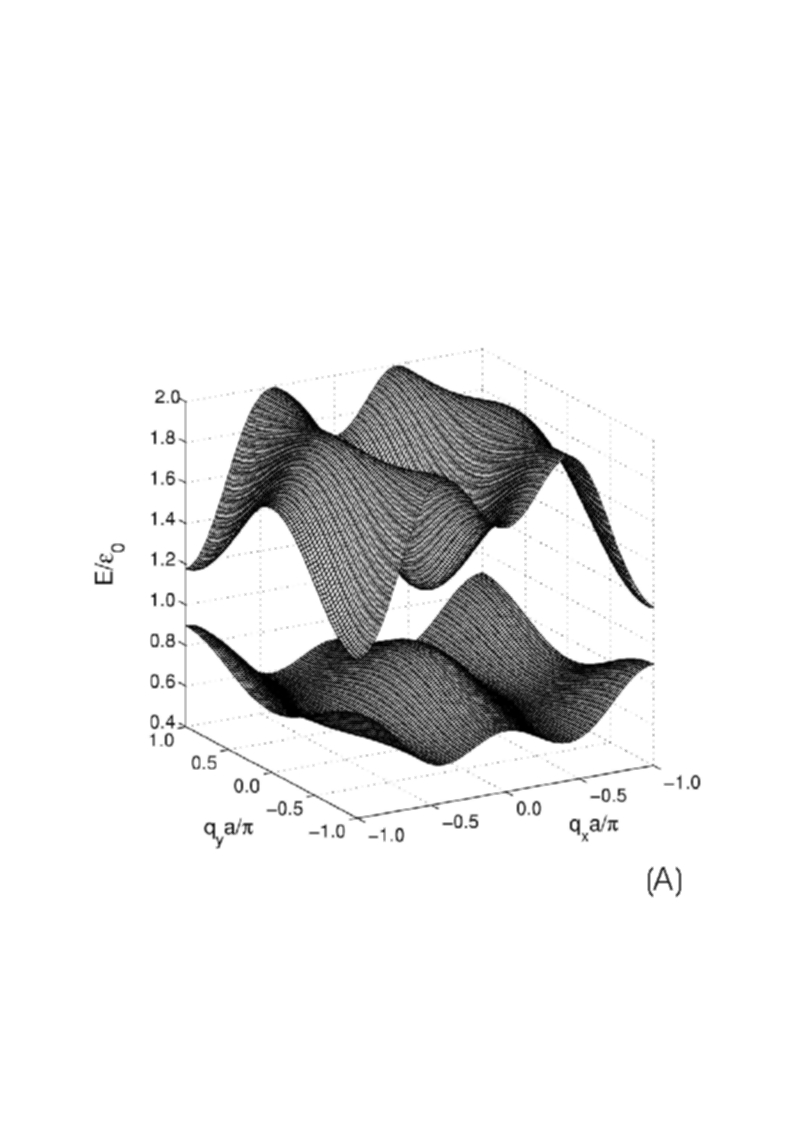}
\includegraphics [width=0.55\textwidth ,bb= 45 45 530 530]{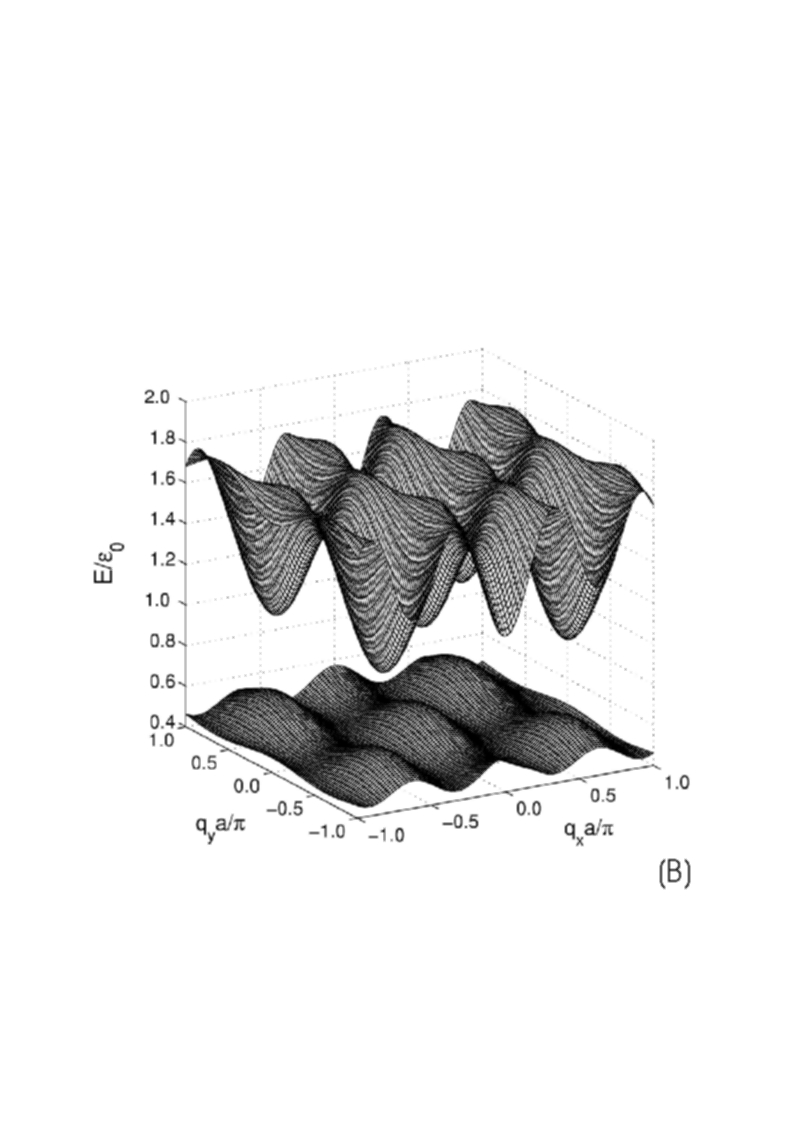}
\caption{ Magnetic energy bands $n=0,1$,of the triangular lattice
as a function of the crystal momentum $(q_x,q_y)$ for or
(A)\hspace{1pt} $\Phi/\Phi_0=2 $. and (B) for $\Phi/\Phi_0=5/4 $.
The modulation strengths are \hspace{2pt} $V_x=V_y=0.5$ meV and
the energy eigenvalues are plotted in units of $\epsilon_0=\hbar
\omega$. }
%%\label{ss3.f1}
\end{figure}

\begin{figure}[htb]
\centerline{\psfig{file=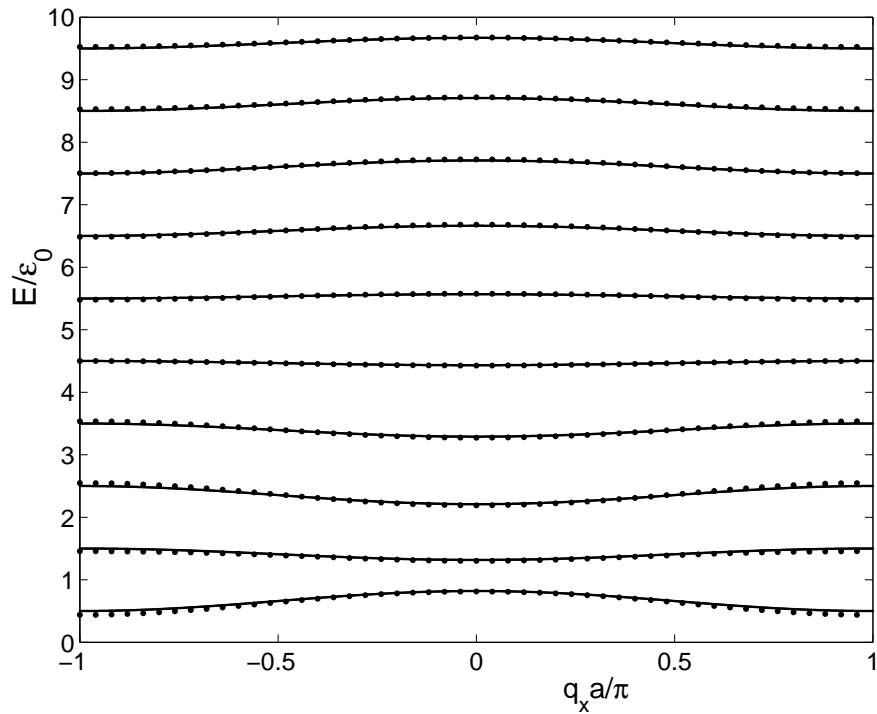,width=4.65in}} \vspace*{8pt}
\caption{Energy spectrum of the square lattice as a function of
the crystal momentum  $(q_x,q_y=0)$  for the same values of
parameters as in Fig.2. Solid lines are for the first order energy
perturbation and dotted lines are for the exact energy spectrum.
Only the lowest ten levels are shown. The energy eigenvalues are
plotted in units of $\epsilon_0=\hbar \omega$}. \label{fig.5}
%\end{center}
\end{figure}

\end{document}